\documentclass[pdflatex,sn-aps]{sn-jnl}%

\usepackage{graphicx}%
\usepackage{multirow}%
\usepackage{amsmath,amssymb,amsfonts}%
\usepackage{amsthm}%
\usepackage{mathrsfs}%
\usepackage[title]{appendix}%
\usepackage{xcolor}%
\usepackage{textcomp}%
\usepackage{manyfoot}%
\usepackage{booktabs}%
\usepackage{algorithm}%
\usepackage{algorithmicx}%
\usepackage{algpseudocode}%
\usepackage{listings}%
\usepackage{ulem}
\usepackage[T1]{fontenc}
\usepackage[version=4]{mhchem}

\definecolor{Mycolor1}{RGB}{0,128,0}

\unnumbered 

\begin{document}

\title{Coherent Spin Waves in Curved Ferromagnetic Nanocaps of a 3D-printed Magnonic Crystal}


\author[1]{\fnm{Huixin}\sur{Guo}}
\email{huixin.guo@epfl.ch}
\equalcont{These authors contributed equally to this work.}

\author*[2]{\fnm{Kilian} \sur{Lenz}}
\email{k.lenz@hzdr.de}
\equalcont{These authors contributed equally to this work.}

\author[3]{\fnm{Mateusz}\sur{Go\lębiewski}}
\email{mateusz.golebiewski@amu.edu.pl}
\equalcont{These authors contributed equally to this work.}

\author[2]{\fnm{Ryszard}\sur{Narkowicz}}
\email{r.narkovic@hzdr.de}

\author[2]{\fnm{J\"urgen}\sur{Lindner}}
\email{j.lindner@hzdr.de}

\author[3]{\fnm{Maciej}\sur{Krawczyk}}
\email{krawczyk@amu.edu.pl} 

\author*[1,4]{\fnm{Dirk}\sur{Grundler}}
\email{dirk.grundler@epfl.ch}

\affil[1]{\orgdiv{School of Engineering, Institute of Materials,
Laboratory of Nanoscale Magnetic Materials and Magnonics}, \orgname{École Polytechnique Fédérale de Lausanne (EPFL)}, \orgaddress{\city{Lausanne} \postcode{1015}, \country{Switzerland}}}

\affil[2]{\orgdiv{Institute of Ion Beam Physics and Materials Research}, \orgname{Helmholtz-Zentrum Dresden--Rossendorf}, \orgaddress{\street{Bautzner Landstr.\ 400}, \postcode{01328} \city{Dresden}, \country{Germany}}}

\affil[3]{\orgdiv{Institute of Spintronics and Quantum Information,
Faculty of Physics and Astronomy}, \orgname{Adam Mickiewicz University}, \orgaddress{\street{Uniwersytetu Poznańskiego 2}, \postcode{61-614} \city{Poznań}, \country{Poland}}}

\affil[4]{\orgdiv{School of Engineering, Institute of Electrical and Micro Engineering}, \orgname{École Polytechnique Fédérale de Lausanne (EPFL)}, \orgaddress{\city{Lausanne} \postcode{1015}, \country{Switzerland}}}


\abstract{
Coherent magnon modes in a truly three-dimensional (3D) magnonic crystal have not yet been investigated. This scientific gap exists despite the numerous theoretical predictions about miniband formation and edge modes with topological protection. Such properties are key to advance nanomagnonics for ultrafast data processing. In this work, we use a scalable nanotechnology and integrate a 3D magnonic crystal to an on-chip microresonator. It was fabricated by two-photon lithography of a 3D woodpile structure and atomic layer deposition of 30-nm-thick nickel. Operated near 14 and 24~GHz, the microresonator output revealed numerous coherent magnons with  distinct angular dependencies reflecting the underlying face-centred cubic lattice. Micromagnetic simulations show that the edge modes are localised in curved nanocaps and robust against changes in field orientation. Along an edge, they exhibit an unexpected phase evolution. Our findings advance functional microwave circuits with 3D magnonic crystals and fuel their visionary prospects of edge-dominated magnon modes.
}
\maketitle





Three-dimensional (3D) nanomagnetic systems have gained significant attention in recent years, serving as a platform for both fundamental discoveries and practical applications \cite{Gub2019,LOS2022,GBL2024, FSF2017,Sheka2022,MVK2022,CGC2022,LPK2023}. The geometry and topology in 3D architectures give rise to new physics \cite{streubel_magnetism_2016, MVK2022, DFG2020,schobitz_fast_2019,Her2013, GOLEBIEWSKI2025}. Meanwhile, their inherent advantages---high storage density, device miniaturisation, and numerous interconnections---make them promising for next-generation applications, including memory, logic, and neuromorphic computing. Among these, 3D magnonic crystals stand out by utilising spin waves instead of charge transport, thereby eliminating Joule heating and making them highly attractive for energy-efficient information processing. Additionally, they offer reprogrammable magnonic band structures, enabling tunable spin-wave propagation for advanced wave-based computing \cite{Gub2019,KrP2008}, including topologically protected edge magnon modes in 3D architectured magnetic materials \cite{Shindou2013}.

Although the concept of 3D racetrack memory was introduced years ago in 2008 \cite{PHT2008}, research on 3D magnetic nanodevices remained mostly theoretical due to fabrication challenges. Only recently have advances in nanofabrication techniques, such as focused electron beam-induced deposition (FEBID) \cite{DHA2022}, nanosphere-lithography \cite{MAlbrecht2005,PhysRevB.85.174429,10.1063/1.5007213}, two-photon lithography (TPL) \cite{DFW2004,LFB2007,MaF2008,MZC2015,HTA2020,SMW2018}, and block copolymer templating \cite{RBC2014}, enabled the realisation of 3D magnetic nanostructures. These developments have now allowed for direct experimental investigations, including spin-wave characterisation and magnetic imaging \cite{DFG2020}. Though the racetrack broke free recently from the substrate, a scalable nanotechnology  is not yet available for a full 3D implementation \cite{PachecoNatNano}.  

Despite selected advances, a systematic study of spin dynamics in 3D magnonic crystals is still missing. Previous investigations of thermally or coherently excited spin waves in 3D magnetic systems have lacked full three-dimensional periodicity. For instance, in \cite{DOVV2021}, FEBID-fabricated, 40-nm-thick Co--Fe nanovolcano structures were studied using spin-wave resonance spectroscopy, revealing that the ring-shaped regions support high-frequency eigenmodes, whereas the crater regions host lower-frequency excitations. Sahoo~\textit{et al.}~\cite{SSM2021} investigated coherent spin-wave modes in a 3D artificial spin-ice lattice using Brillouin light scattering (BLS), with simulations predicting both localised and extended modes. Still, the system consisted of only four magnetic sublayers. The absence of several periods in vertical direction precludes the classification of previously investigated samples as 3D magnonic crystals. More recently \cite{GMH2024}, gyroid-based ferromagnetic nanostructures have been explored for their collective spin-wave dynamics experimentally and numerically. However, due to the presence of multiple crystallographic domains in a large volumetric sample, the measured spectra only provided an averaged response.

To address these limitations, the authors of Ref.~\cite{GDX2023} developed a scalable technology for creating true 3D magnonic crystals by combining TPL with atomic layer deposition (ALD) of ferromagnetic nickel (Ni). Using micro-focused BLS and exploring incoherent magnons excited in 3D ferromagnetic woodpile structures by thermal fluctuations, they identified surface modes of ultrahigh frequency with shifts of up to 10~GHz between surface and bulk modes. As the BLS was inherently surface-sensitive, the magnon modes deep within the bulk of the samples and on their side facets remained inaccessible. Integration to on-chip microwave devices and the coherent excitation of spin waves in 3D magnonic crystals have yet to be established.

\begin{figure*}[t]
\centering
\includegraphics[width=0.95\textwidth]{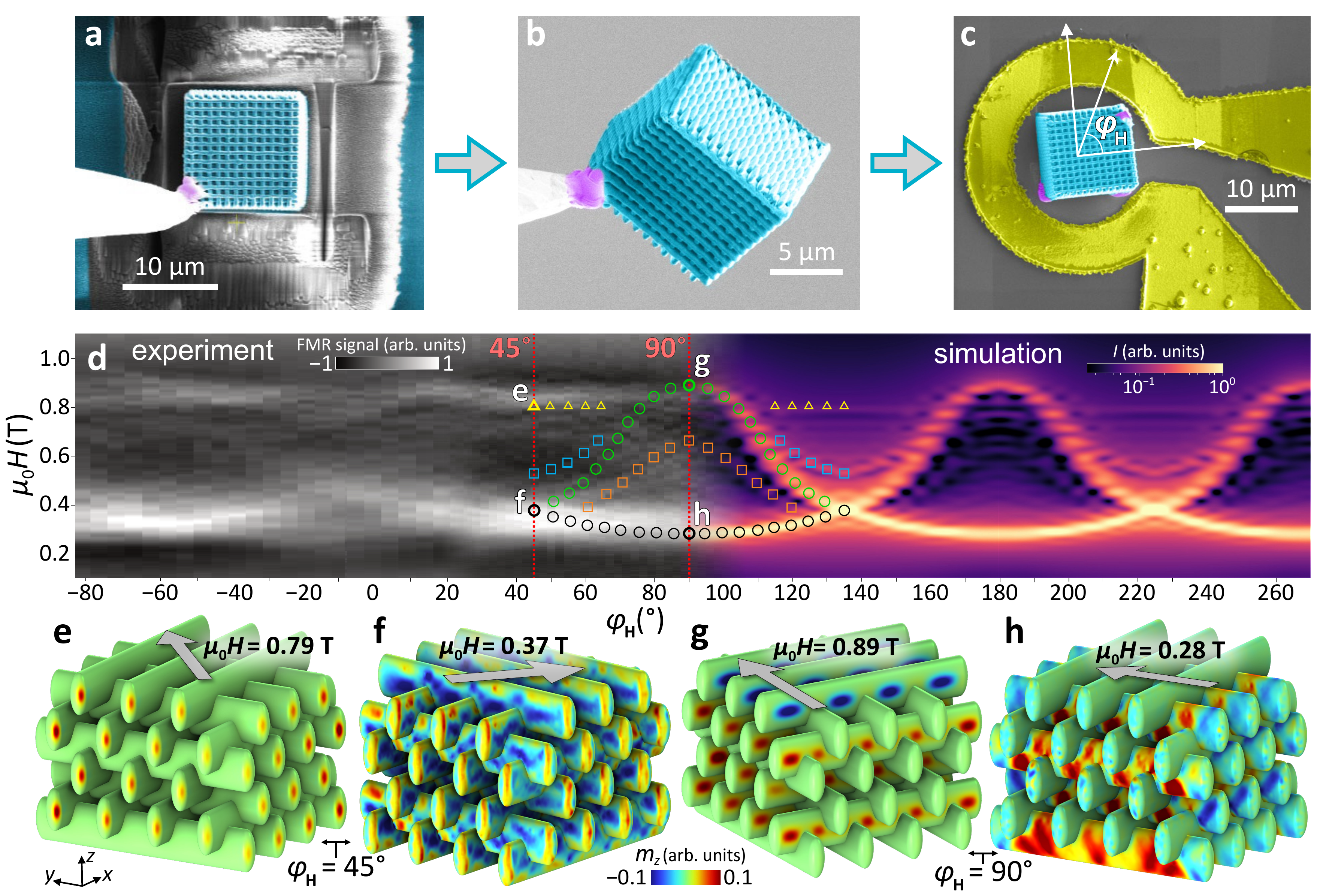}
\caption{\textbf{Woodpile preparation, ferromagnetic resonance measurements, and micromagnetic simulations.} \textbf{a}, False-coloured scanning electron microscopy (SEM) image of the Ni-coated (blue) woodpile structure on the Si substrate (grey). \textbf{b}, Released woodpile structure attached to the micromanipulator. Platinum (Pt, purple) was deposited to attach the micromanipulator. \textbf{c}, The woodpile structure mounted inside the planar microresonator (yellow) for ferromagnetic resonance measurements with the coordinate system of the in-plane magnetic field angle $\varphi_H$. For $\varphi_H = 0^\circ$ the external field \textbf{H} is parallel to the top-row of the Ni nanotubes in the woodpile. \textbf{d}, The angular dependence of the resonance field at 14.26~GHz reveals a multitude of spin-wave modes, captured experimentally (greyscale, left) and through micromagnetic simulations (colour scale, right). Symbols highlight characteristic magnon mode branches as visual guides. Letters denote resonances whose spatial distributions of the normalised dynamic magnetisation $z$-component are shown in \textbf{e--h}.}
\label{Fig:showcase}
\end{figure*}
In this work, we report an essential advancement towards the development of functional 3D magnonic devices by integrating a full-fledged 3D magnonic crystal into a microwave resonator. Its thin-film micro-loop antenna (Figs.~\ref{Fig:showcase}\textbf{a--c}) is operated at about 14 and 24~GHz. The comparison between the experimental spectra and micromagnetic simulations (Fig.~\ref{Fig:showcase}\textbf{d}) reveals that the high sensitivity of the microresonator enables the detection of both bulk and edge-localised modes. The analysis uncovers several classes of spin-wave excitations, including extended modes propagating along both principal directions of the nanotubes, as well as spatially confined modes localised at specific structural regions, such as the conformally coated end-caps of the woodpile lattice. For some of these localised modes their eigenfrequencies are found to be strikingly independent of the field orientation over wide field and angular ranges. These results represent a significant advance toward the coherent excitation and control of spin waves in three-dimensional magnonic crystals. They also establish a viable route for integrating 3D nanomagnetic architectures into functional microwave circuits through inductive coupling, highlighting their potential for reconfigurable and miniaturised magnonic devices.

\subsection*{Nanofabrication, experiment and simulations}\label{Nano}

\begin{figure*}[t]
\centering
\includegraphics[width=0.95\textwidth]{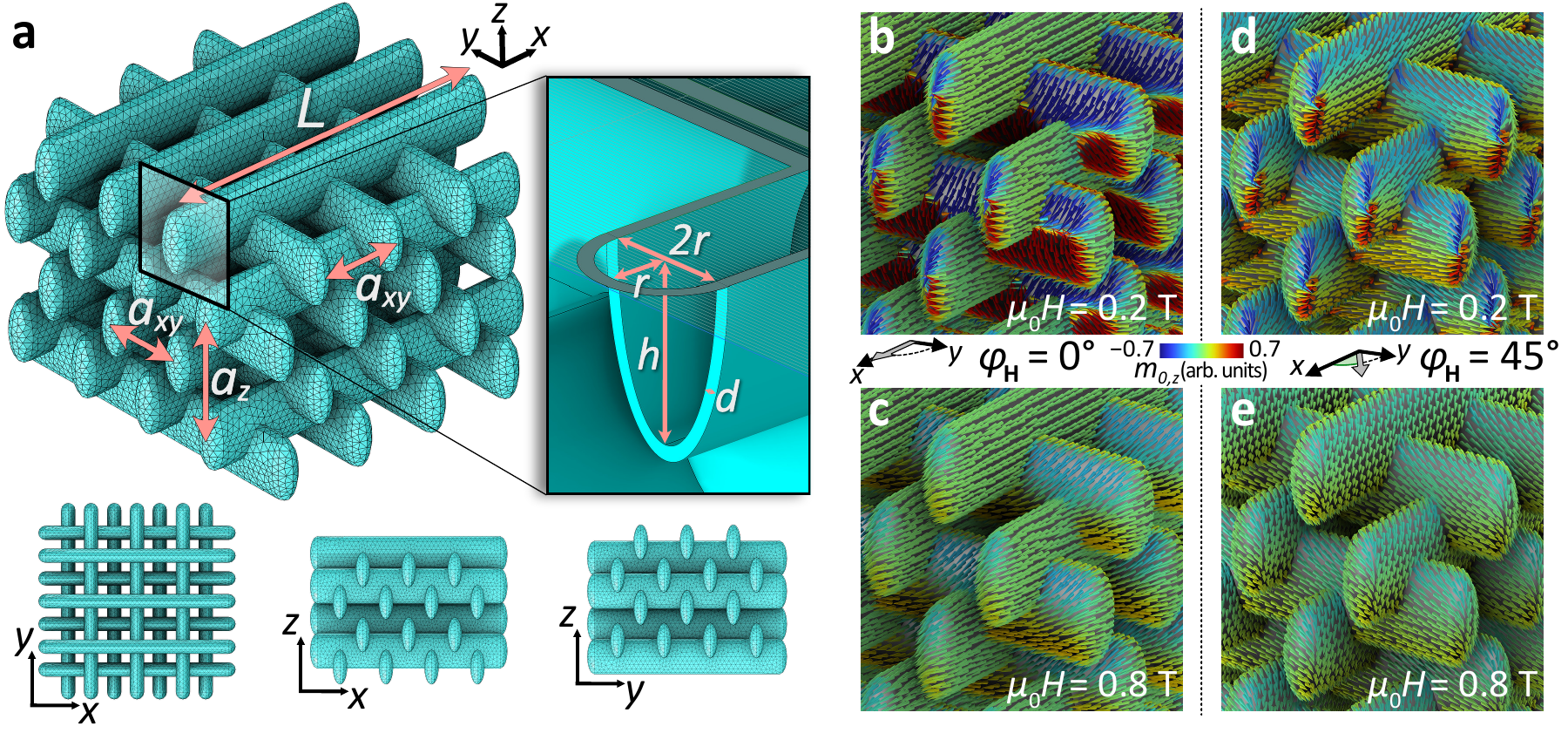}
\caption{\textbf{FEM mesh and simulated static magnetisation configuration.} \textbf{a}, FEM mesh of the woodpile model used in the \textsc{Comsol} micromagnetic simulations. The structure consists of eight sublayers of periodically stacked magnetic nanotubes, arranged with a horizontal spacing of $a_{xy} = 1$~\textmu m and a vertical spacing of $a_{z} = \sqrt{2}a_{xy} \approx 1.41$~\textmu m. Each has a length of $L = 4.31$~\textmu m and an elliptical cross-section with semi-axes $h = 350$~nm and $r = 125$~nm, respectively. The ends are capped with rounded terminations also of depth $r$, as shown in the inset. The thickness $d$ is assumed to be uniform throughout the structure and set to $d = 30$~nm. \textbf{b--e}, Static magnetisation configurations for selected in-plane field values and field angles: \textbf{b,c}, along the $x$-axis ($\varphi_H=0^\circ$) and \textbf{d,e}, rotated by $45^\circ$. The cones represent the local equilibrium direction of the magnetisation $\mathbf{m}_0$ after relaxation, while the colour indicates its normalised $z$-component $m_{0,z}$.}
\label{Fig:SimDimensions}
\end{figure*}

The magnonic crystals were fabricated through the scalable additive manufacturing process, combining TPL of a polymeric nanotemplate with Ni-ALD as described in the methods section and Refs.~\cite{GBE2020,GDX2023}. Using this 3D-printing methodology, we fabricated several micrometre high woodpiles consisting of $12\times12\times6$ unit cells of ferromagnetic nanotubes arranged in a face-centred cubic (fcc) lattice. The outer dimensions of the woodpile were $11.7\times 11.7\times 8.4$~\textmu m$^3$, and the in-plane lattice constant was $a_{xy} = 1$~\textmu m. The polymeric nanotemplate was conformally ALD-coated with a 5-nm-thick \ce{Al2O3} layer, serving as an insulating and adhesion-promoting interface, followed by a 30-nm-thick Ni layer deposited using nickelocene as the metalorganic precursor. The elliptical cross-section of one of the Ni tubes measures 700~nm $\times$ 250~nm. The nanotubes are terminated by ellipsoidal ferromagnetic end-caps, characterised by a semi-major axis $h=350$~nm and a semi-minor axis $r=125$~nm, as illustrated in the geometric model in Fig.~\ref{Fig:SimDimensions}\textbf{a}. 

Figure~\ref{Fig:showcase}\textbf{a--c} illustrates the integration of the 8.4-\textmu m-high woodpile nanostructure into a planar microresonator fabricated on a separate substrate. The process begins with Xe plasma focused ion beam (FIB) etching (\textbf{a}), which detaches the woodpile from its original silicon support. The structure is then attached to a micromanipulator needle (\textbf{b}) and fully released for transfer. The purple regions denote the deposited Pt used as a mechanical stabiliser (`glue'), while blue indicates the conformal ALD-grown Ni coating on both the 3D architecture and the silicon substrate. The panel \textbf{c} finally shows the woodpile positioned within the microresonator coil, along with the coordinate system used to define the orientation of the in-plane magnetic field $\mathbf{H}$. Here, the angle $\varphi_H$ specifies the direction of $\mathbf{H}$ relative to the long axis of the Ni nanotubes in the topmost layer, such that $\varphi_H = 0^\circ$ corresponds to alignment along the top row of nanotubes.

Ferromagnetic resonance (FMR) measurements were performed using a home-built spectrometer employing a bridge-type detection scheme in reflection mode \cite{LNW2019,PKH2023}. We applied a microwave frequency of 14.26~GHz to the resonator antenna and explored the microwave absorption of the 3D Ni structure via inductive coupling. This coupling is significantly enhanced due to the large filling factor inside the loop of the planar resonator, thus boosting its detection sensitivity far beyond conventional cavities. In the centre of the loop the RF magnetic field is pointing along the $z$-axis (normal to the plane) and perpendicular to the long axes of the Ni nanotubes forming the 3D woodpile. 
The FMR signals were taken as a function of the external in-plane field $\mathbf{H}$ applied at different angles $\varphi_{H}$. Figure \ref{Fig:showcase}\textbf{d} shows the angle-dependent FMR data as a greyscale intensity plot, i.e., resonances appear bright.

In order to gain microscopic insight into the woodpile's magnetisation dynamics and to identify the various resonant modes, we simulated the three-dimensional nanostructure using the finite-element method (FEM) software package \textsc{Comsol Multiphysics} \cite{comsolv61}. This approach allows us to account for spatially varying magnetic fields throughout the complex geometry of the woodpile sample.

Full 3D micromagnetic simulations place significant demands on the computational resources of the state-of-the-art infrastructure used in this work. For the woodpile model, we therefore consider 8 vertically stacked layers of nanotubes (i.e., two woodpile-type unit cells) and 7 layers nanotubes stacked along the horizontal direction. The tetrahedral FEM mesh for the simulations is depicted in Fig.~\ref{Fig:SimDimensions}\textbf{a}. The mesh-size varies depending on the local curvature and near the junctions. Despite the reduced total number of nanotubes compared to the real sample, the vertical stacking symmetry and the overall lattice geometry are maintained to ensure consistency with the experimentally realised structure. Figures~\ref{Fig:SimDimensions}\textbf{b--e} show the simulated static spin configurations in the Ni shell at magnetic fields of 0.2 and 0.8~T applied at angles $\varphi_H$ of zero and $45^\circ$, respectively. These field values are below and above the maximum shape anisotropy field in Ni, which amounts to $\mu_0M_{\rm s}=0.6~$T. Accordingly, the non-collinear spin structures in the tubular segments and curved nanocap regions, such as onion states and vortex cores \cite{Sheka2022,10.1063/1.5007213}, are prominent for the small field, but transform into an (almost fully) aligned/saturated configuration at the large field value. The fundamental dimensions like thickness of the Ni shell, tubular cross-section, top-down symmetry, and lattice periods agree with the investigated sample. Our static and dynamic simulations hence consider the essential magnetostatic and dynamic properties of the experimentally studied woodpile structure, enabling a direct comparison with the measured FMR spectra.

\subsection*{Extended and localised 3D magnon modes at 14.26~GHz}\label{secFMRResults}
The left-hand side of Fig.~\ref{Fig:showcase}\textbf{d} presents the experimental FMR spectra obtained at $f = 14.26$~GHz as a greyscale map plotted as a function of the in-plane field angle $\varphi_H$. For each angle, the external magnetic field was swept from saturation ($\mu_0 H = 1.2$~T) down to zero. Due to the use of lock-in amplification with field modulation, the measured signal corresponds to the field-derivative of the FMR response, in contrast to the Lorentzian-like susceptibility spectra obtained from micromagnetic simulations. To enable direct comparison, we plot the real part of the derivative, $\Re(\partial \chi / \partial H)$, which captures the symmetric, peak-like features of the resonant absorption. In this representation, white regions indicate strong spin-precessional absorption in the 3D Ni structure.

The spectra reveal multiple resonance branches with varying signal strengths. Some exhibit a pronounced angular dependence of their resonance field $H_{\rm res}$ with a near-$90^\circ$ periodicity, while others remain nearly angle-independent. A cluster of strong resonances appears around 0.3~T for most angles between $-80^\circ$ and $80^\circ$. Another prominent group emerges at higher fields, between 0.8 and 0.9~T. These regions are interconnected by several weaker branches, which cross near 0.6~T at $\varphi_H \approx 0^\circ$ and $90^\circ$, exhibiting both positive and negative slopes $dH_{\rm res}/d\varphi_H$. Convincingly, the coherent driving of the 3D magnonic crystal by a microresonator gives rise to many more and sharper magnon branches than preliminary experiments in which incoherent magnons on the top surface were explored by inelastic light scattering over a broad regime of wave vectors $k$ \cite{GDX2023}.
 
Before discussing selected resonance branches and their associated spin-precessional dynamics, it is instructive to first perform a quantitative comparison between the experimentally measured and simulated resonance fields. Figure~\ref{Fig:showcase}\textbf{d} contains a superposition of the experimental data (left, greyscale) and the simulated data (right), overlaid with symbols marking selected mode branches, which we extracted from the simulations. The simulated branches cover the same field range as the measured ones, i.e.\ from about 0.3 to 0.9 T. Upon closer examination---and in agreement with our experimental observations---the simulated spin-wave modes can be broadly classified into two categories: (i) flat branches, which are largely angle-independent (indicated by triangles), and (ii) curved branches, which exhibit pronounced angular variation (marked by circles and squares). The good quantitative agreement between the simulated and measured resonance fields, along with the consistent angular trends, provides a solid foundation for analysing the corresponding spin-precessional mode profiles and assigning them to the observed spectral features.

\begin{figure}[t]
\centering
\includegraphics[width=0.7\columnwidth]{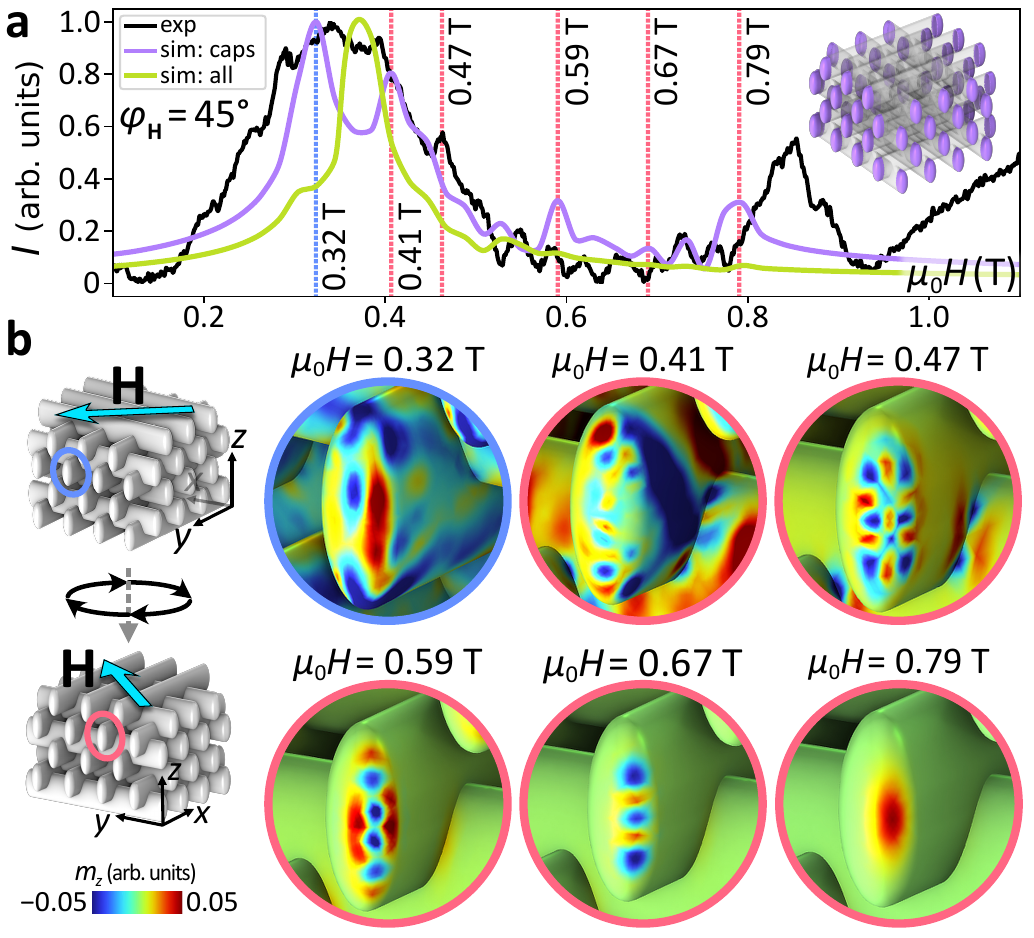}
\caption{\textbf{Localised modes at the tube caps.} \textbf{a}, FMR spectra at $\varphi_H=45^\circ$ and excitation frequency $f=14.26$~GHz, obtained from both simulation and experiment. The green curve corresponds to simulated intensities integrated over the entire woodpile structure, see Eq.~(\ref{eq:intens}), while the purple curve reflects integration restricted to the cap regions (highlighted in the inset, ${\sim}255$~nm inward from each edge). The experimental spectrum recorded at the same angle is shown in black. Each curve in panel \textbf{a} is independently normalised to its own maximum intensity; therefore, no direct comparison of absolute amplitudes between simulation and experiment is implied. \textbf{b}, Simulated spin-wave mode profiles in the woodpile structure for selected field values. Each panel displays the spatial distribution of the normalised out-of-plane dynamic magnetisation component, $m_z$, highlighting the localization and geometry of cap modes.}
\label{Fig:SimCaps14GHz}
\end{figure}

The flat branches include multiple resonances at different magnetic field values and are observed near the yellow triangles in both the experimental and simulated spectra, particularly around $\varphi_H = 40^\circ$ and $130^\circ$, respectively. These modes exhibit minimal sensitivity to changes in the field angle, which is consistent with their localised character. Micromagnetic simulations confirm that the high-field modes are indeed confined to the cap regions of the structure (Fig.~\ref{Fig:showcase}\textbf{e}). For $\varphi_H = 45^\circ$, the simulations reveal an entire family of cap-localised spin-wave modes (Fig.~\ref{Fig:SimCaps14GHz}; see also Extended Data Fig.~\ref{Fig:SimCaps14GHz_90deg} for results at $\varphi_H = 90^\circ$).
Between 0.47 and 0.79~T, four localised modes with relatively high intensity are identified, as shown in Fig.~\ref{Fig:SimCaps14GHz}\textbf{b}. As the external field increases, the wavelength of these modes increases (wave vector $k$ decreases) in order to maintain the same resonance frequency of 14.26~GHz---indicating a decrease in mode energy with increasing field. At 0.79~T, the spin-wave profile becomes strongly confined, exhibiting nearly uniform, spot-like localisation in neighbouring caps (Fig.~\ref{Fig:showcase}\textbf{e}). In contrast, at 0.47~T, the simulated spin-precessional profile reveals a more complex mode with a nontrivial spatial interference pattern and dynamic localisation at the nanotube terminations (Fig.~\ref{Fig:SimCaps14GHz}\textbf{b}). For fields below 0.47~T, the precessional motion extends deeper into the bulk of the woodpile structure. As shown in Fig.~\ref{Fig:showcase}\textbf{f}, the highest dynamic intensities are then observed in the flat segments of the tubes, and the mode is no longer localised under a small applied field at $\varphi_H = 45^\circ$.

The pronounced resonances observed at $\varphi_H = 0^\circ$, $90^\circ$, and $180^\circ$ in Fig.~\ref{Fig:showcase}\textbf{d} occur when the external magnetic field is aligned along the longitudinal axis of one half of the nanotubes---that is, along either the $x$- or $y$-axis in the simulation coordinate system. Depending on whether the resonance field is high or low, the corresponding modes are spatially excited in different regions, as illustrated in Figs.~\ref{Fig:showcase}\textbf{g} and \textbf{h}, respectively. For lower $\mu_0 H_{\rm res}$, the modes reside predominantly within the bulk of the nanotubes that are aligned with the applied field. The reduced resonance field in this regime is consistent with magnetisation aligned along an easy axis, as described in Ref.~\cite{Gurevich}. Supporting this interpretation, a comparable mode and $H_{\rm res}$ were also identified in simulations of a single Ni nanotube subjected to a longitudinal magnetic field $\mathbf{H}$ (see Supplementary Figs.~S1\textbf{c--f}). Naturally, in contrast to the isolated nanotube, the modes in the 3D woodpile structure are more complex due to the intersections of orthogonal nanotube segments and the dipolar coupling to neighbouring tubes in the lattice. Consequently, the uniform excitation results in characteristic chequerboard-like interference patterns induced by the scattering of the coherently excited spin waves at intersections and tube ends. At a large field of 0.89~T (Fig.~\ref{Fig:showcase}\textbf{g}), the resonant spin precession is mainly in nanotube segments which are perpendicular to the applied field, i.e., $\mathbf{H}$ points into the hard-axis direction. The excitation is concentrated in the region exhibiting the strongest demagnetising field  \cite{CGC2022}. This occurs because, at higher fields, the static magnetisation aligns nearly parallel to the applied field (see Fig.~\ref{Fig:SimDimensions}\textbf{c}), and the demagnetising field peaks in regions where the local surface normal is parallel to the magnetisation vector $\mathbf{M}$. Modes \textbf{g} and \textbf{h} are, in fact, bulk excitations. At $\varphi_H = 45^\circ$, they become degenerate and merge into mode \textbf{f}, as expected from symmetry considerations. Consequently, all modes observed at $\varphi_H = 45^\circ$ and at fields exceeding the degeneracy point (0.37~T) must originate from the edge regions, specifically the nanocaps. This distinction is explored in more detail in the following section.

\subsection*{Phase-coherent spin precession in curved nanocaps}\label{secCapModeResults}

Notably, the conformally coated woodpile structures support a distinct class of high-field spin-wave modes that are spatially confined to the curved end-caps of the nanotubes, as shown in Fig.~\ref{Fig:showcase}\textbf{e}. These resonant modes are localised in regions where the demagnetising field is strongest. As $\mu_0 H$ increases beyond 0.47~T, the mode becomes increasingly confined (Fig.~\ref{Fig:SimCaps14GHz}\textbf{b}). Owing to their spatial confinement to a limited volume, their contribution to the simulated FMR spectrum is diminished when averaged over all cells. Consequently, the corresponding spectrum (green curve in Fig.~\ref{Fig:SimCaps14GHz}\textbf{a}) exhibits a markedly narrower linewidth for the peak between 0.35 and 0.45~T compared to the experimental spectrum (black curve), which instead appears as a superposition of multiple overlapping peaks spanning 0.25 to 0.5~T. 

To account for these side peaks, we generated a separate spectrum from the micromagnetic simulations by isolating the dynamic contributions of the tube ends (purple curve in Fig.~\ref{Fig:SimCaps14GHz}\textbf{a}). The integration volume in Eq.~(\ref{eq:intens}) was restricted to the nanotube end-cap regions on both sides of the structure, as indicated in purple in the inset of Fig.~\ref{Fig:SimCaps14GHz}\textbf{a}. This targeted approach selectively captures the resonances arising from the caps, revealing two distinct modes in the low-field regime at 0.32 and 0.41~T. A corresponding cap-only analysis was also conducted for $\varphi_H = 90^\circ$ (see Extended Data Fig.~\ref{Fig:SimCaps14GHz_90deg}).

In the experimental spectrum (black curve in Fig.~\ref{Fig:SimCaps14GHz}\textbf{a}), there is a further pronounced resonance peak visible in the high-field regime between 0.8 and 0.9~T. This feature is absent in the spectrum accumulated over the full woodpile structure (green curve). Interestingly, while this mode is barely visible in the full-structure simulation, it appears as a pronounced peak at 0.79~T in the cap-only spectrum, closely matching the experimental signal. Modes in Fig.~\ref{Fig:SimCaps14GHz}\textbf{b} appearing below 0.79~T correspond to higher-order cap excitations, producing multiple local maxima in the purple curve. Overall, the experimental spectrum aligns more closely with the cap-only simulation than with the full-volume one. 

The detection of cap modes, despite being relatively weak in the simulation, highlights the excellent sensitivity of our microresonator-based FMR detection scheme. We attribute this to the spatial distribution of the microwave magnetic field within the coil (see Extended Data Fig.~\ref{Fig:ResonatorSim}), which plays a crucial role in mode excitation and detection. While the simulations assume uniform excitation across the entire 3D structure, the experimentally generated magnetic field from the micro-loop is approximately 2.5 times stronger at the edges and corners than in the central region. This non-uniform field distribution provides a compelling explanation for the enhanced visibility of cap-localised modes in experiment.
\begin{figure}[t]
\centering
\includegraphics[width=0.95\columnwidth]{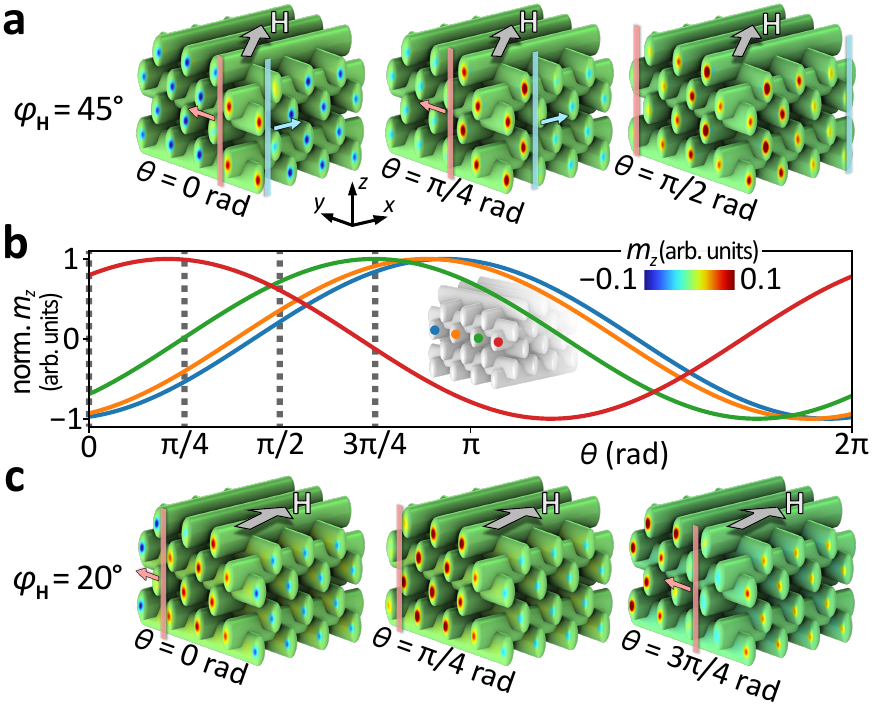}
\caption{\textbf{Phase-dependent spin-wave dynamics of cap modes.} Cap-localised spin-wave modes exhibit wave-like propagation across the side facets of the 3D woodpile structure, visualised here for two selected in-plane field orientations and dynamic phases at $f=14.26$~GHz. \textbf{a}, Time-resolved snapshots of the out-of-plane dynamic magnetisation component $m_z$ at three distinct phases, $\theta = 0$, $\pi/4$, and $\pi/2$~rad of the exciting microwave field, for an applied magnetic field of 0.77~T oriented at $\varphi_\mathrm{H} = 45^\circ$. The observed periodic pattern reflects coherent propagation of the cap mode across the structure. \textbf{b}, Phase-dependent visualisation of the normalised magnetisation component $m_z$ in selected caps (highlighted in the inset), with colours linking each reference cap to its corresponding curve. \textbf{c}, Same representation as in \textbf{a}, but for a reduced field angle of $\varphi_H = 20^\circ$ and phases $\theta = 0$, $\pi/4$, and $3\pi/4$~rad. Together, the panels demonstrate the dynamic character and spatial coherence of cap-localised spin-wave modes, underscoring their potential for collective behaviour and field-tunable phase control.}
\label{Fig:cap_waves}
\end{figure}
It is important to note that we do not directly compare absolute mode intensities between simulation and experiment. The fundamentally different excitation mechanisms and selection rules---uniform field excitation in simulations versus spatially inhomogeneous microwave fields in the experiment---preclude a quantitative correspondence in mode amplitudes. Rather, the experimental spectrum can be viewed as a weighted superposition of responses from different regions of the structure, effectively combining features captured in both the full-structure and the cap-only spectrum. This interplay is particularly evident in the spectra shown in Fig.~\ref{Fig:SimCaps14GHz}\textbf{a}, where distinct contributions from both the extended (bulk) and cap-localised modes collectively shape the measured signal (black curve).

To probe the coherence of the cap-localised modes, we examine whether they precess in phase, as one might expect under the uniformly applied microwave field assumed in the simulations. Figure~\ref{Fig:cap_waves}\textbf{a} shows the spin-precessional amplitudes extracted from the dynamic simulations at three points in time for $f = 14.26$~GHz and $\varphi_H = 45^\circ$. Surprisingly, the different caps reach their maximum spin-precessional amplitudes at different times (Fig.~\ref{Fig:cap_waves}\textbf{b}), indicating that the excitation is not uniformly phased across the structure. The four caps analysed in Fig.~\ref{Fig:cap_waves}\textbf{b} are positioned along a central row, as shown in the inset, with colours corresponding to the respective amplitude traces. It is important to emphasise that these plots show normalised magnetisation amplitudes---each trace is individually scaled to its own maximum. As a result, the curves convey only the relative phase evolution, without reflecting differences in absolute amplitude between the caps. Nevertheless, they clearly reveal that the dynamic magnetisation in the caps exhibits a wave-like phase gradient across the structure along its lateral edges. Vertical lines in Fig.~\ref{Fig:cap_waves}\textbf{a} serve as visual guides, illustrating the temporal evolution of the phase across two adjacent side facets. Animations of this are included in the Supplementary Material. 

This counterintuitive behaviour is also observed at $\varphi_H = 20^\circ$ (Fig.~\ref{Fig:cap_waves}\textbf{c}) for the same magnetic field value of $0.77$~T. Similar to the case of $\varphi_H = 45^\circ$, the spins in the caps exhibit a continuous phase shift across the surface of the woodpile structure, and their amplitudes follow the same normalised profiles as shown in Fig.\ref{Fig:cap_waves}\textbf{b}. In the $\varphi_H = 20^\circ$ field configuration, the wave-like character of the phase front becomes even more pronounced, particularly on the side of the structure oriented more perpendicular to the applied magnetic field $\mathbf{H}$. 

Our findings underscore a key result: the curved nanocaps support collective, edge-localised spin-wave modes with coherent, tunable phase relationships. The spatial coherence across neighbouring caps is clearly resolved, revealing a propagating phase gradient that emerges despite the uniform microwave excitation. This striking observation might reflect an alternative realisation of the unusual nature of confined modes in chiral magnetic systems \cite{Zingsem2019}. This emergent phase ordering, arising within a three-dimensional architecture under spatially uniform excitation, offers a promising platform for phase-based information processing and directional signal routing in 3D magnonic systems.

\subsection*{Resonance fields and their magnon modes beyond shape anisotropy fields}\label{secBeyond}
\begin{figure*}[t]
\centering
\includegraphics[width=0.95\textwidth]{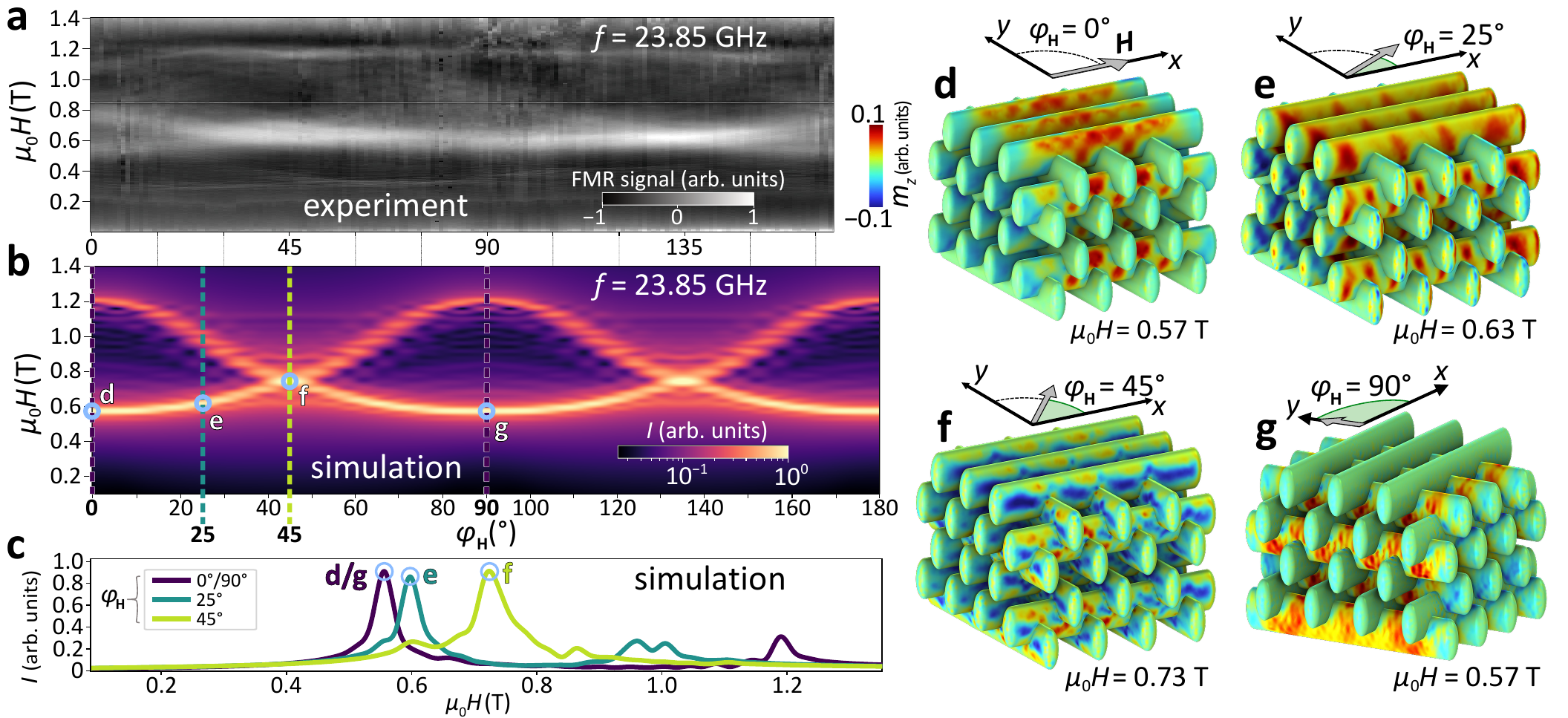}
\caption{\textbf{FMR experiment and simulations at 23.85~GHz.} \textbf{a}, Experimentally measured absorption spectra as a function of magnetic field strength, recorded at various in-plane field orientations $\varphi_H$ for an excitation frequency of 23.85~GHz. The field was swept from 1.4~T down to zero, with white colour corresponding to maximum absorption. \textbf{b}, Simulated angular-resolved spectra reproduce the key features observed in the experiment. Dashed lines and letters mark selected field angles, for which panel \textbf{c} shows the corresponding simulated FMR spectra, enabling clearer identification of mode branches. \textbf{d--g}, The spatial profiles of the normalised out-of-plane dynamic magnetisation component $m_z$ for the labelled resonances illustrate the diversity of spin-wave character in the 3D woodpile structure. For $\varphi_H = 0^\circ$, the applied field is aligned parallel to the top row of the Ni nanotubes.}
\label{Fig:Sim24GHz_spec}
\end{figure*}

The minimum resonance field observed in the experiments presented so far is approximately 0.3~T, which lies below the maximum shape anisotropy field of the ALD-grown polycrystalline Ni, estimated to be 0.5~T. To understand the influence of shape anisotropy fields, we performed additional measurements at a much higher frequency of 23.85~GHz. Although our microresonator was optimised for a microwave resonance at 14.26~GHz, we were still able to operate it off-resonantly. For these measurements, a Mach--Zehnder-type microwave interferometer was added to the signal path of the FMR spectrometer. This reduced the very high idle off-resonant reflection of the microresonator by more than 35~dB; thus, we avoided saturation of the low-noise amplifier. The greyscale graph in Fig.~\ref{Fig:Sim24GHz_spec}\textbf{a} combines the spectra taken from high to small magnetic fields at different angles $\varphi_H$.
As expected, the resonances observed at 23.85~GHz appear at higher magnetic fields $\mu_0 H_{\rm res}$—above 0.5~T due to the increased excitation frequency. The most prominent features are centred around 0.6~T and 1.2~T, with several weaker branches connecting them across the intermediate field range. Ni nanowires studied at almost the same frequency in Ref. \cite{Ebels2001} showed only two branches for which $H_{\rm res}$ followed identical 180-degree rotational symmetries. 

While the resonance branches that we observe at 14.26 and 23.85~GHz share overall similarities---including a consistent shift of resonance fields with increasing frequency---a difference arises in weaker spin-precessional amplitudes consistent with the higher-frequency excitation, where the magnetic susceptibility is reduced and  the number of prominently resolved modes at $\varphi_H = 0^\circ$ in the experimental data increased, as seen by comparing Fig.~\ref{Fig:showcase}\textbf{d} and Fig.~\ref{Fig:Sim24GHz_spec}\textbf{a}. Four distinct branches are clearly visible near 0.55, 0.75, 1.1 and 1.2~T. We attribute this improved mode resolution and the narrow linewidths to the larger applied field, which facilitates saturated magnetic states across all branches. Interestingly, the authors of the previous study on incoherent magnons \cite{GDX2023} identified their richest spectral characteristics and most well-separated modes also when the magnetic field was aligned with the long axis of the top-layer nanotubes. By spatially resolved spectroscopy, three distinct modes were observed at zero angle in Fig.~4 of that work. The three modes were attributed to the centre-, tube-end-, and corner-regions of the top surface of the 3D woodpile structure, respectively. 


The simulations shown in Fig.~\ref{Fig:Sim24GHz_spec}\textbf{b} represent power spectra integrated over the entire 3D woodpile volume. Three representative line cuts are displayed in Fig.~\ref{Fig:Sim24GHz_spec}\textbf{c}.  At $\varphi_H = 0^\circ$, the simulations predict fewer prominently resolved branches compared to our experimental results. In the dark violet curve of Fig.~\ref{Fig:Sim24GHz_spec}\textbf{c}, only two clear peaks are observed. This discrepancy can be attributed to the spatial averaging inherent to the simulation approach, which dilutes the contribution of modes that are strongly localised in the cap regions.
Along the low-field branches, the spatial mode profiles (Figs.~\ref{Fig:Sim24GHz_spec}\textbf{d--g}) closely resemble those found at 14.26~GHz (Figs.~\ref{Fig:showcase}\textbf{e--h}), indicating that the fundamental nature of the excitations remains largely unchanged. As before, we observe a coexistence of bulk-like and edge-localised modes, governed by field orientation and geometric confinement (see Sec.~S3 of the Supplementary Material). Despite the reduced amplitude, the cap dynamics at 23.85~GHz reveal a more pronounced wave-like behaviour. This is attributed to the shorter effective wavelength (i.e., larger wave vector) at higher frequencies, which results in faster spatial oscillations and more distinct phase fronts across the caps---see Sec.~S3 of the Supplementary Material. As a consequence, phase coherence between neighbouring caps becomes more readily visible, reinforcing the interpretation of these modes as coherent, propagating edge excitations.

Taken together, these findings support the interpretation that microresonators exhibit excellent sensitivity to edge-localised magnon modes---modes that are of particular interest in the context of 3D topological magnonics.

\section*{Conclusions}\label{secConcl}
In summary, we have successfully integrated a full-fledged 3D magnonic crystal into a microresonator, enabling the coherent excitation and detection of its magnon modes. The angular dependence of the spectra at the resonator frequency of 14.26~GHz revealed a clear fourfold symmetry (90-degree rotational symmetry), consistent with the in-plane geometry of the woodpile lattice. We experimentally observed a rich set of spin-wave modes, reflecting the complex unit cell of the fcc 3D structure. Our micromagnetic simulations based on the same geometry showed a good quantitative agreement with the measured resonance fields at both 14.26 and 23.85~GHz. 

By visualising the simulated mode profiles, we identified both extended modes spanning the full 3D network and localised modes confined to the curved nanocaps. These edge-localised modes, resolved in both simulation and experiment, showed up in flat branches. They exhibited robustness against variations in the magnetic field orientation and display a directed phase evolution under uniform excitation---intriguing features that may suggest topological characteristics.

Interestingly, the microresonator loop exhibited enhanced sensitivity to such edge magnon modes. This work establishes a versatile platform for studying coherent spin dynamics in 3D magnonic systems and highlights the potential for further engineering of their band structure. It also demonstrates a practical approach to integrating complex 3D magnetic architectures into functional microwave devices.

\section*{Online content}
Any methods, additional references, Nature Portfolio reporting summaries, source data, extended data, supplementary information,
acknowledgements, peer review information; details of author contributions and competing interests; and statements of data and code availability are available at https://doi.org/10.1038/...



\backmatter

\section*{Methods}

\subsection*{Sample fabrication}\label{secSample}

We fabricated the 3D polymer nanoscaffold for the woodpile by using a two-photon lithography system (Nanoscribe Photonic Professional GT+) at the Center of MicroNanoTechnology at EPFL. A droplet of IP-Dip2 photoresist (refractive index $n=1.547$) was deposited onto a silicon substrate ($10\times 10$~mm$^2$, 700~\textmu m thick). We used the dip-in laser lithography configuration with a 63$\times$ magnifying objective and infrared femtosecond laser (780~nm) for localised polymerisation in PiezoScan mode. The laser power was adjusted to 10~mW during writing. Following the exposure, the substrate underwent a sequential development process by initial immersion in propylene glycol monomethyl ether acetate for 20 minutes to remove unexposed resist. Then the structures were rinsed for 5 minutes in isopropyl alcohol and dried in a nitrogen flow. Finally, the substrate, along with additional bare silicon reference substrates, was placed inside the ALD chamber of a Beneq TFS200 system at CMi. We first deposited a 5-nm-thick \ce{Al2O3} layer, followed by a 30-nm-thick ferromagnetic Ni layer using nickelocene as a precursor. The details of the ALD deposition process are described in Refs.~\cite{GBE2020,GDX2023}. The choice of silicon as substrate was intentional to minimise charging effects during the subsequent FIB etching processes.

To transfer the woodpile into the microresonator, we used Xe-plasma FIB etching by means of a FEI Helios G4 plasma-FIB UXe system at the Interdisciplinary Centre for Electron Microscopy at EPFL. It combines a FIB with a SEM. To facilitate the detachment of the woodpile structure from the Si substrate, three trenches were defined by FIB milling on three adjacent sides of the structure. A micromanipulator needle was then attached to the top corner of the woodpile by depositing a Pt patch (see Fig.~\ref{Fig:showcase}\textbf{a}). To remove the Si substrate from the base of the 3D structure, the fourth side was subsequently cut at the bottom of the woodpile. We employed a 30~kV ion beam with currents ranging from 15~nA to 1~nA, depending on the proximity to the structure’s base. After positioning the nanostructure inside the omega-shaped thin-film microresonator coil we secured it by spot-deposition of Pt at two bottom corners and cut off the manipulator needle by Xe FIB.

\subsection*{Microresonator ferromagnetic resonance}

The planar microresonator design was optimised for an eigenfrequency of about 14.26~GHz and 50~$\Omega$ impedance using the \textsc{Ansys HFSS} simulation package \cite{NSS2005,NSN2008}. The nominal inner loop diameter was 20~\textmu m. It was prepared on a highly resistive Si(001) substrate by means of photolithography, molecular-beam epitaxy of Cr/Cu/Au, and subsequent lift-off processing \cite{BNH2011,LNW2019,NaS2015}. The metallic loop antenna had a total thickness of about 700~nm. In the centre of the loop the RF magnetic field was along the $z$-axis (out of plane) and perpendicular to the long axes of the Ni nanotubes forming the 3D woodpile. Simulations of the RF-field distribution are given in Extended Data Fig.~\ref{Fig:ResonatorSim}.

The FMR measurements were performed using a home-built spectrometer in reflection mode \cite{LNW2019,PKH2023}. The signals were taken as a function of the in-plane field $\mathbf{H}$ applied at different angles $\varphi_H$. An additional field-modulation with 1~mT amplitude at 78~kHz was employed to allow for lock-in detection. 

The measurements at 23.85~GHz are far off the eigenfrequency of the resonator. This results in the idle resonator reflecting approximately 20~dB more power, which would lead to saturation of the low-noise amplifier (LNA) circuit. To mitigate this issue, we employed an interferometric approach that enabled cancellation of the idle signal prior to entering the LNA. This was achieved by splitting the microwave signal and introducing a $\pi$ phase shift along with careful attenuation to one of the paths. The two signals were then recombined using the Mach--Zehnder interferometer principle, effectively cancelling more than 35~dB of the background trace and allowing mainly the resonance signal to be passed to the LNA and detector. This avoids saturating the low-noise pre-amplifier. Thereby, measurements with a large signal-to-noise ratio far off the resonator's eigenfrequency become possible. The microwave power at the sample was set to $-10$~dBm~$=100$~\textmu W without the interferometer, and to $+4$~dBm with the interferometer. Using an IQ-mixer (in-phase and quadrature mixer) as detector in conjunction with two lock-in amplifier channels we retrieved simultaneously the field derivatives of the FMR absorption $\partial\chi^{\prime\prime}/\partial H$ and dispersion $\partial\chi^{\prime}/\partial H$ signals.

\subsection*{Micromagnetic simulations}

We used the FEM-based \textsc{Comsol Multiphysics} software package \cite{comsolv61} for our micromagnetic simulations. We employ the \textit{Coefficient Form} interface, a flexible framework that enables the definition and solution of custom partial differential equations. In this study, it is used to implement the Landau--Lifshitz--Gilbert equation [Eq.~(S1) of the Supplementary Material]. We used the following material parameters for Ni matching our experimental conditions (see Extended Data Fig.~2): saturation magnetisation $M_\mathrm{s}=400$~kA/m as reported for the ALD-grown Ni \cite{GBE2020}, exchange stiffness $A_\mathrm{ex}=8$~pJ/m, and the gyromagnetic ratio $\gamma=193.6$~GHz/T. During each time-domain (relaxation) simulation, the Gilbert damping constant $\alpha$ is set to a high value to ensure rapid convergence to the equilibrium state. For the subsequent frequency-domain simulations, it is reduced to $\alpha = 0.008$. Further methodological details are provided in the Supplementary Information and in Refs.~\cite{GOLEBIEWSKI2025,Mruczkiewicz2013StandingCrystals,Rychy2018SpinRegime}.

To emulate the FMR condition, we apply a spatially uniform dynamic magnetic field of magnitude $\mu_0 h_\mathrm{RF} = 1$~mT, polarised along the $z$-axis (always perpendicular to $\mathbf{H}$). By sweeping the static field magnitude and keeping the microwave frequency $f$ fixed, the magnetisation response throughout the volume is computed. Then, to highlight the global resonance signature, we integrate the complex, normalised dynamic magnetisation vector, $\mathbf{m} = (m_x, m_y, m_z) = \mathbf{M}/M_\mathrm{s}$. In particular, we focus on its $m_z$ component, which is oriented perpendicular to the static magnetic field. The resonance intensity is defined as
\begin{equation}
    I = \left(\int_V \Re\{m_z\}\,\mathrm{d}V\right)^2
      + \left(\int_V \Im\{m_z\}\,\mathrm{d}V\right)^2.
      \label{eq:intens}
\end{equation}
This integral captures the overall resonance intensity of the sample by combining the real and imaginary parts of $m_z$ integrated over the entire sample volume $V$. In selected cases, the integration is restricted to the cap regions, defined as the side-edge sections of the woodpile structure extending approximately 255~nm inward from each edge. An unstructured tetrahedral mesh is generated throughout the ferromagnet and its surrounding region, with finer elements concentrated within and around the Ni shell to accurately resolve the magnetic interactions. The complete 3D model of the woodpile structure (excluding the surrounding region) as shown in Fig.~\ref{Fig:SimDimensions}\textbf{a} consists of approximately 150\,000 elements.

\section*{Data availability}
The data of this work are included in the published article and its Supplementary Information. All other data supporting the findings of this study are available from the corresponding author upon reasonable request.

\section*{Code availability}
The simulation codes are available from the corresponding author upon reasonable request.

\section*{Acknowledgements}

This work was supported by the SNSF via grant number 197360. This research used the nanofabrication facilities in the Center of Mi\-cro\-Nano\-Tech\-no\-lo\-gy (CMi) and the Interdisciplinary Center for Electron Microscopy (CIME) at EPFL. We acknowledge technical support by D. Bouvet. Support by the Nanofabrication Facilities Rossendorf (NanoFaRo) at HZDR is gratefully acknowledged. This work was also supported by the National Science Centre, Poland (UMO-2020/39/I/ST3/02413 and UMO-2023/49/N/ST3/03032). M.G. received funding from the Adam Mickiewicz University Foundation in Poznań (2024/2025).

\section*{Author contributions}
H.G., K.L., and M.G. contributed equally to this work and wrote the paper with inputs from all authors. H.G., K.L., and D.G. conceived the work. H.G. prepared the samples. K.L. performed and analysed the FMR measurements. M.G. and M.K. performed and interpreted the numerical simulations. R.N. designed, simulated, and prepared the microresonators. D.G., M.K. and J.L. supervised the project. All authors read and commented on the paper.

\section*{Competing interests}
The authors declare no competing interests.

\section*{Additional information}

\noindent\textbf{Extended data} is available for this paper at https://doi.org/10.1038/...\\

\noindent\textbf{Supplementary information} The online version contains supplementary material available at https://doi.org/10.1038/...\\

\noindent\textbf{Correspondence and requests for materials} should be addressed to Dirk Grundler or Kilian Lenz. \\

\bmhead{Peer review information} \textit{Nature Nanotechnology} thanks the anonymous reviewers for their contributions to the peer review of this work.\\

\noindent\textbf{Reprints and permissions information} is available at www.nature.com/reprints.


\begin{appendices}
\label{secA1}

\renewcommand{\figurename}{Extended Data Fig.}


\begin{figure}[t]
\centering
\includegraphics[width=0.95\columnwidth]{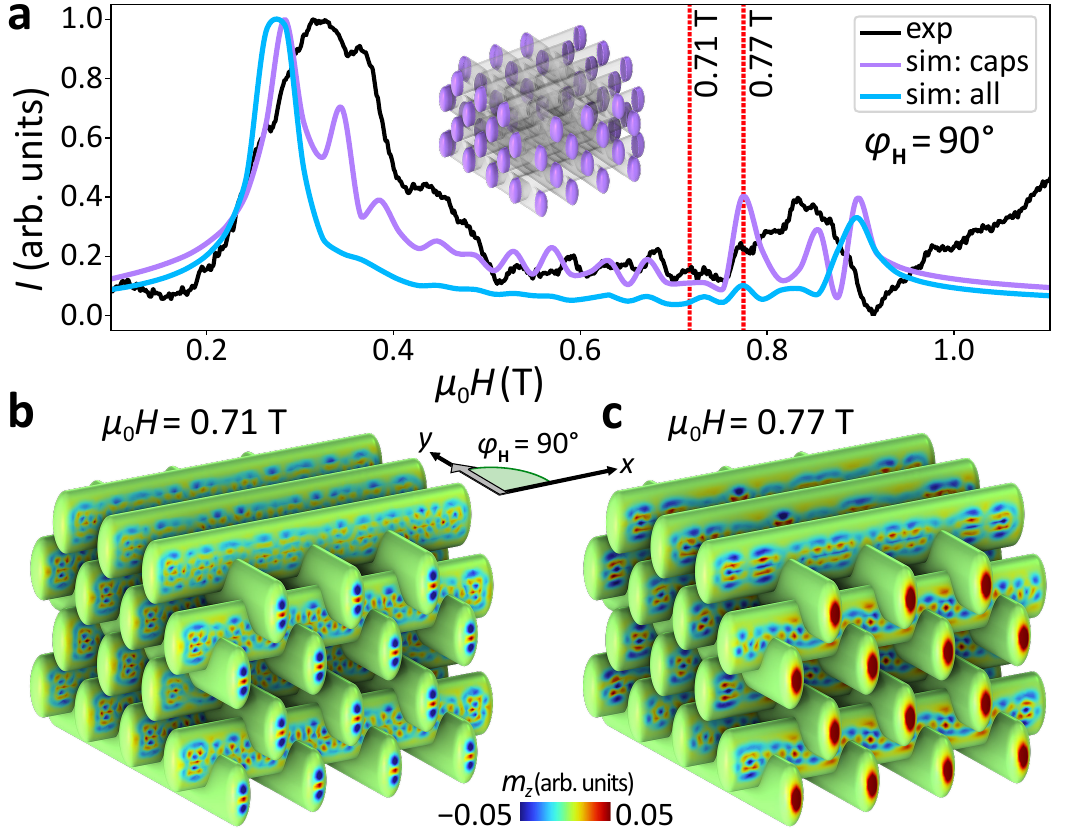}
\caption{\textbf{Comparison of cap modes at $\varphi_{\mathrm{H}} = 90^\circ$.} Analogous to the $\varphi_{\mathrm{H}} = 45^\circ$ case shown in Fig.~\ref{Fig:SimCaps14GHz}, this figure presents results for $\varphi_{\mathrm{H}} = 90^\circ$ and excitation frequency $f = 14.26$~GHz, obtained from both simulation and experiment. In \textbf{a}, the blue curve represents simulated intensities integrated over the entire woodpile structure, see Eq.~(\ref{eq:intens}), while the purple curve corresponds to integration limited to the cap regions, as indicated in the inset. The experimental spectrum recorded at the same field angle is shown in black. Each curve is independently normalised to its own maximum, and thus no direct comparison of absolute amplitudes is implied. \textbf{b}, Simulated spin-wave mode profiles for selected magnetic field values, showing the spatial distribution of the normalised out-of-plane dynamic magnetisation component, $m_z$. The images highlight the localisation and symmetry of cap modes in the woodpile geometry.}
\label{Fig:SimCaps14GHz_90deg}
\end{figure}

\begin{figure}[t]
\centering
\includegraphics[width=0.75\columnwidth]{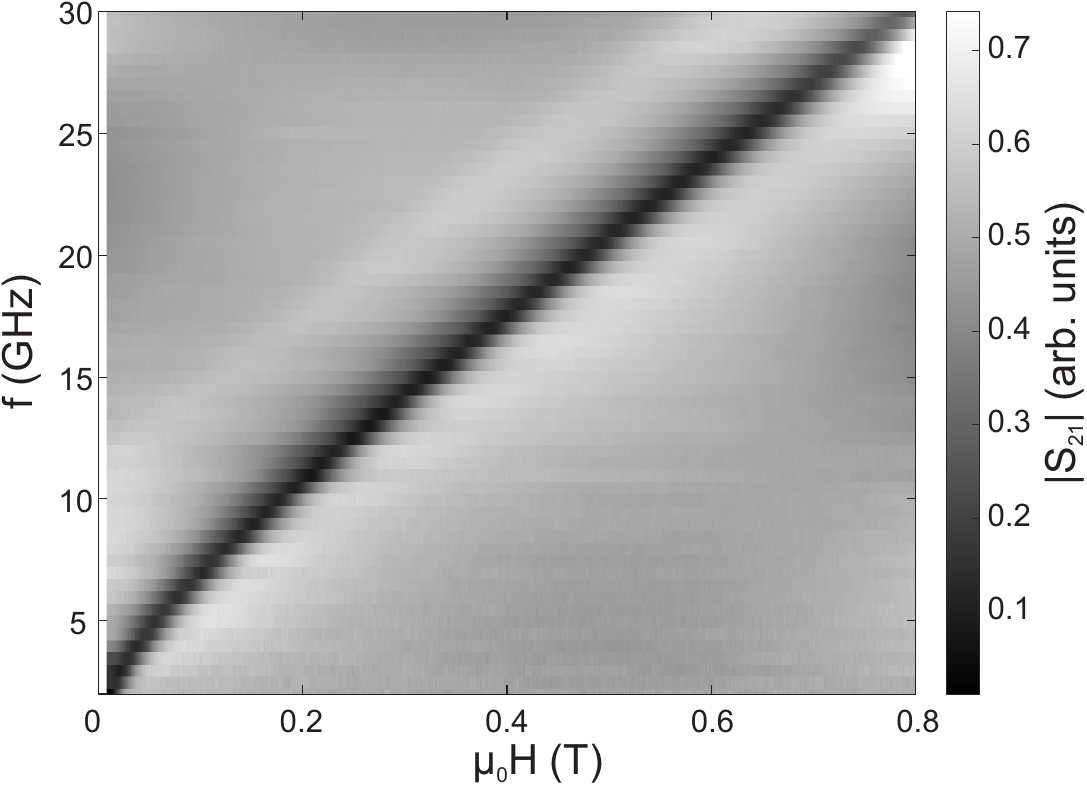}
\caption{\textbf{FMR measurement of the Ni-ALD reference sample.} Intensity plot of the magnitude of the scattering parameter $S_{21}$ showing the in-plane frequency-field dependence measured by vector-network analyser FMR on the Ni-ALD reference thin film. We determined an effective magnetisation of $\mu_0M_\mathrm{eff}=400$~mT, and a $g$-factor of $g=2.174$ in strong consistency with previous reports \cite{GBE2020, GDX2023}, confirming the reproducibility and robustness of our approach. Analysis of the linewidth yields a Gilbert damping of $\alpha=0.025$, which is slightly lower than what was found for thin Ni/Si(001) films \cite{WKD2008}.}
\label{Fig:FMR_Ni}
\end{figure}

\begin{figure*}[t]
\centering
\includegraphics[width=0.95\textwidth]{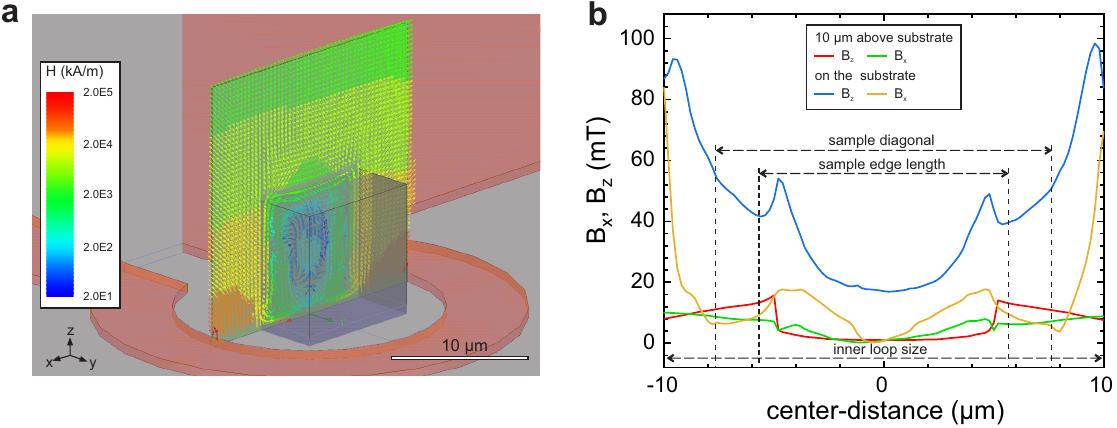}
\caption{\textbf{Simulated RF-field distribution of the microresonator antenna.} \textbf{a}, Field distribution within the microresonator loop antenna with respect to the sample (at 14.26~GHz and 0~dBm power) simulated using the \textsc{Ansys HFSS}. \textbf{b}, Line profiles of the $B_x$ and $B_z$ field strength at the bottom and top of the sample. The arrows mark the sample and antenna dimensions. As the resonator antenna is only 700~nm high, the bottom of the 11.7~\textmu m large sample experiences the strongest RF excitation field of about 20-40~mT compared to the top. The RF field is even stronger at the corner regions of the structure, as they are the closest part to the antenna.}
\label{Fig:ResonatorSim}
\end{figure*}

\end{appendices}

\end{document}